\documentclass[a4paper,oneside,11pt]{article}
\usepackage{myf}
\usepackage{amsmath,amstext,amsfonts,amsbsy,amssymb,amscd,bbm,epsfig,lscape}

\newcommand{\ba}{\begin{array}}
\newcommand{\ea}{\end{array}}

\newcommand{\rep}[1]{\cite{#1}}

\newcommand{\dif}{{\rm d}}

\newcommand{\Dslash}{\relax{\kern+.25em / \kern-.70em D}}

\newcommand{\Real}{\relax{\mathsf{\Gamma\kern-.35em R}}}
\newcommand{\Int}{\relax{\mathsf{Z\kern-.40em Z}}}

\newcommand{\SUT}{\mbox{SU}(2)}

\newcommand{\SUt}{\mbox{SU}(3)}

\newcommand{\gbar}{\kern1pt\overline{\kern-1pt g\kern-0pt}\kern1pt}
\newcommand{\mbar}{\kern2pt\overline{\kern-1pt m\kern-1pt}\kern1pt}
\newcommand{\obar}[1]{\kern3pt\overline{\kern-2pt #1\kern-0pt}\kern1pt}

\newcommand{\mrgi}{M}

\newcommand{\msubt}{m_{\rm q}}

\newcommand{\hopc}{\kappa_{\rm c}}

\newcommand{\fX}{f_{\rm\scriptscriptstyle X}}
\newcommand{\fP}{f_{\rm\scriptscriptstyle P}}

\newcommand{\fA}{f_{\rm\scriptscriptstyle A}}

\newcommand{\ZP}{Z_{\rm\scriptscriptstyle P}}

\newcommand{\ZA}{Z_{\rm\scriptscriptstyle A}}

\newcommand{\sigmaP}{\sigma_{\rm\scriptscriptstyle P}}

\newcommand{\SigmaP}{\Sigma_{\rm\scriptscriptstyle P}}

\newcommand{\lmax}{L_{\rm max}}

\newcommand{\Oa}{\mbox{O}(a)}
\newcommand{\Oasq}{\mbox{O}(a^2)}
\newcommand{\Ogsqa}{\mbox{O}(g_0^2 a)}
\newcommand{\Ogqa}{\mbox{O}(g_0^4 a)}
\newcommand{\icsw}{c_{\rm sw}}
\newcommand{\ict}{c_{\rm t}}
\newcommand{\icttil}{\tilde c_{\rm t}}

\newcommand{\icA}{c_{\rm\scriptscriptstyle A}}

\newcommand{\abar}{\kern1pt\overline{\kern-1pt a\kern-0pt}\kern1pt}

\newcommand{\vy}{\mathbf{y}}
\newcommand{\vz}{\mathbf{z}}

\begin{document}


\begin{titlepage}


\vspace*{-30truemm}
\begin{flushright}
ROM2F/2004-04\\
MS-TP-04-02\\
DESY 04-022\\
\end{flushright}
\vspace{15truemm}


\centerline{\Bigrm The continuum limit of the quark mass}
\vskip 2 true mm
\centerline{\Bigrm step scaling function in quenched lattice QCD}
\vskip 9 true mm
\begin{center}
\epsfig{figure=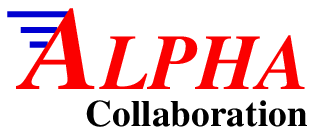, width=22 true mm}\\
\end{center}
\vskip -2 true mm
\centerline{\bigrm  M.~Guagnelli$^a$, J.~Heitger$^b$, F.~Palombi$^{c,a}$, C.~Pena$^d$ and A.~Vladikas$^a$}
\vskip 4 true mm
\centerline{\it $^a$ INFN, Sezione di Roma II}
\centerline{\it c/o Dipartimento di Fisica, Universit\`a di Roma ``Tor
  Vergata''}
\centerline{\it Via della Ricerca Scientifica 1, I-00133 Rome, Italy}
\vskip 3 true mm
\centerline{\it $^b$ Westf\"alische Wilhelms-Universit\"at M\"unster,
Institut f\"ur Theoretische Physik}
\centerline{\it Wilhelm-Klemm-Strasse 9, D-48149 M\"unster, Germany}
\vskip 3 true mm
\centerline{\it $^c$ ``E. Fermi'' Research Center}
\centerline{\it c/o Compendio Viminale, pal. F, I-00184 Rome, Italy}
\vskip 3 true mm
\centerline{\it $^d$ DESY, Theory Group}
\centerline{\it Notkestrasse 85, D-22607 Hamburg,
  Germany}
\vskip 10 true mm


\thintablerule
\vskip 3 true mm
\noindent{\tenbf Abstract}
\vskip 1 true mm
\noindent
{\tenrm The renormalisation group running of the quark mass is
determined non-perturbatively for a large range of scales, by
computing the step scaling function in the Schr\"odinger Functional
formalism of quenched lattice QCD both with and without $\Oa$
improvement. A one-loop perturbative calculation of the discretisation
effects has been carried out for both the Wilson and the
Clover-improved actions and for a large number of lattice
resolutions. The non-perturbative computation yields continuum
results which are regularisation independent, thus providing
convincing evidence for the uniqueness of the continuum limit. 
As a byproduct, the ratio of the renormalisation group invariant quark
mass to the quark mass, renormalised at a hadronic scale, is obtained with
very high accuracy.}
\vskip 3 true mm
\thintablerule
\vspace{10truemm}

\eject
\end{titlepage}


\section{Introduction}
\label{sec:intro}

The renormalisation group running of the QCD fundamental parameters,
namely the renormalised gauge coupling and quark masses,
has now been computed non-perturbatively for a large range of scales,
albeit in the limit of infinitely heavy sea quarks; see
refs.~\cite{SFcoupling,SFmassRGI}. These results have been obtained
using lattice regularised quenched QCD with Wilson fermions, prior to 
extrapolating to the continuum limit. In the case of the quark mass running,
Symanzik improvement was an important element in reducing the 
extrapolation uncertainties, as it implies that the dominant systematic 
effects due to the finiteness of the UV cutoff $a^{-1}$ are O$(a^2)$.
The case of QCD with two dynamical flavours has been investigated
in~\rep{alpha_nf2}.

The continuum limit of lattice QCD is known to exist to all orders of
perturbation theory (PT) \cite{Reisz}. Beyond PT this issue has been
addressed by numerical simulation. The strategy consists in implementing
different regularisations which formally correspond to the same Field
Theory (QCD in our case) and in establishing the universality of the continuum
limit of given renormalised physical quantities, computed with different
regulators. This sometimes turns out to be less straightforward
than expected; e.g. see the discussion on the universality of the continuum
limit of spin and sigma models in refs.~\cite{berber}. In pure
$\SUT$ gauge theory, universality has been tested by computing two different
non-perturbatively defined running couplings over a large range of energies~\cite{SFuniv}.
In pure $\SUt$ gauge theory, evidence of universality has been recently
found in a study of the scaling properties of the deconfining temperature
with different gauge actions~\cite{necco}.

In the present work we extend these ideas to the step scaling function (SSF)
of the quark mass in quenched QCD. This quantity has been calculated
in ref.~\cite{SFmassRGI} from an improved action; here it is also evaluated
from an unimproved action. The continuum SSF,
computed for a large range of renormalisation scales, is found to
be independent of these regularisation details, providing evidence
for a universal continuum limit. In this respect our study
parallels closely the one of ref.~\cite{GJP}, which was dealing with the
SSF of the operator corresponding to the average momentum of
non-singlet parton densities. In comparison, our quantity is
particularly simple, as we essentially compute ratios of two-point functions.
This allows us to have an excellent control of both statistical and systematic
errors. As a byproduct we recalculate the flavour independent ratio of the
renormalisation group invariant (RGI) quark mass to the renormalised one (at
a given hadronic scale). We obtain a  result compatible to the original one of
ref.~\cite{SFmassRGI} but fairly more accurate.

A study of discretisation effects in the SSF has also been performed, for both
Wilson and Clover actions, in one-loop PT. This calculation
has been carried out for a large number of lattice resolutions. We find that
lowest order perturbation theory greatly underestimates the discretisation
effects of the SSF.

\section{The Schr\"odinger Functional and $\Oa$ improvement}
\label{sec:SF}

In this section we gather the most relevant definitions and outline
the properties of the quantities we are interested in. Most details
are omitted, as they have been presented in previous works, which we
will frequently refer to.

We adopt the lattice Schr\"odinger functional (SF) formalism~\rep{SF,SFLNWW,SFS};
more specifically we regularise QCD on a lattice of extension $L^3 \times
T$ (here $T=L$ always) with periodic boundary conditions in the space
directions (up to a phase $\theta$ for the fermion fields) and Dirichlet
boundary conditions in the Euclidean time direction~\rep{SFLNWW,SFS}.
Otherwise the lattice gauge and fermionic field actions are of the standard Wilson
type; their $\Oa$ improved version is discussed below. The bare
gauge coupling and quark mass are denoted by $g_0$ (with $\beta \equiv
6/g_0^2$) and $m_0$ (with $2 \kappa \equiv [am_0 + 4]^{-1}$),
respectively.  As we will be working in the quenched approximation,
the bare gauge coupling $g_0$ and chiral point $\hopc$ are
functions of the lattice spacing $a$ alone. The chiral point is
the value of the hopping parameter $\kappa$ for which the
``current'' quark mass, defined below, vanishes.
The bare subtracted quark mass
is defined as $a \msubt = [1/\kappa- 1/\hopc]/2$, whereas an
unrenormalised ``current'' quark mass is given by
\begin{gather}
\label{eq:masscurr}
m(g_0) = \frac{\frac{1}{2}(\partial^\ast_0 + \partial_0) \, \fA(x_0)}{2 \, \fP(x_0)} \,\, ,
\end{gather}
with $\fX~({\rm X}={\rm A},{\rm P})$ the correlation functions of
local bilinear operators
\begin{gather}
\fX(x_0) = -\frac{a^6}{2} \sum_{\vy,\vz} \langle \bar \zeta_j(\vy) \gamma_5
\zeta_i(\vz) \,\, \bar \psi_i(x) \gamma_{\mbox{\tiny X}} \psi_j(x) \rangle \,\, .
\label{eq:2pcorr}
\end{gather}
The field indices $i,j$ label two distinct flavours; the ``boundary fields'' $\zeta$ are defined in
ref.~\cite{SFchiral}.
For ${\rm X}={\rm A}$ we have $\gamma_{\mbox{\tiny X}} = \gamma_0 \gamma_5$ and
for ${\rm X}={\rm P}$ we have $\gamma_{\mbox{\tiny X}} = \gamma_5$.
The forward and backward lattice time derivatives are denoted by
$\partial_0$ and $\partial^\ast_0$ respectively\footnote{We follow closely the notation
of~\rep{SFmassRGI,SFchiral}, whither we refer for details.}. We also
define the correlation function of boundary fields
\begin{gather}
f_1 = -\frac{a^{12}}{2 L^6} \sum_{\vy , \vz , \vy^\prime , \vz^\prime} \,\,
\langle \, \bar \zeta_i^\prime(\vy^\prime) \gamma_5 \zeta_j^\prime(\vz^\prime) \,\,
\bar \zeta_j(\vy) \gamma_5 \zeta_i(\vz) \, \rangle \,\, .
\label{eq:btob}
\end{gather}
Unprimed quantities are defined on the the $x_0=0$ boundary, primed ones 
on the  $x_0=T$ one.

The $\Oa$ Symanzik improvement of the above construction has been
worked out in refs.~\cite{SFLNWW,SFchiral}.  For the pure gauge
action, it amounts to modifying it by introducing time-boundary
counterterms proportional to $[\ict(g_0^2)-1]$. For the fermionic
action we must introduce the well-known clover counterterm in the
lattice bulk, proportional to  $\icsw(g_0^2)$, and time-boundary
counterterms proportional to $[\icttil(g_0^2)-1]$. Correlation
functions of composite operators such as eq.~(\ref{eq:2pcorr}) may
then also be $\Oa$ improved by including in the lattice definition of these
operators the appropriate counterterms. In the chiral limit there are
no such counterterms for the pseudoscalar density $P(x)$, while the
axial current $A_0(x)$ requires the addition of $\partial_0 P$ with a
coefficient $\icA(g_0^2)$. {The axial current is used in the computation
of the bare quark mass, but, being scale independent, it
is clearly not needed in the computation of its renormalisation group running.
Thus $\icA$ will play no r$\hat{\rm  o}$le in the present work.}

All these improvement coefficients may in principle be computed
non-per\-tur\-ba\-tively for a range of values of the bare coupling
$g_0$; for $\icsw$ we rely on the calculation of ref.~\cite{SFcSW}. 
It has also been calculated in perturbation theory to one loop~\cite{Wohlert,SFctt}.
The coefficients $\ict$ and $\icttil$ are known
only in perturbation theory,  to NLO~\cite{SFct} and
LO~\cite{SFctt} respectively:
\begin{align}
\label{eq:ctPT}
\ict(g_0^2) &= 1 - 0.089 g_0^2 - 0.030 g_0^4 \,\, ,
\\
\label{eq:cttPT}
\icttil(g_0^2) &= 1 - 0.018 g_0^2 \, .
\end{align}

A more detailed discussion of perturbative $\Oa$ improvement will
be presented in Section~\ref{sec:ssfPT}. Here we outline the main
expectations related to cutoff effects, in the spirit of the Symanzik
improvement programme~\cite{Symanz}: in the absence of improvement
counterterms (i.e. $\icsw = 0$ and $\ict = \icttil =1$),
correlation functions (such as $\fP$ and $\fA$), computed at fixed UV
cutoff $a^{-1}$ and renormalised non-perturbatively, should exhibit
$\Oa$ deviations from their continuum limit. If all improvement
coefficients were known non-perturbatively, the discretisation errors
would be $\Oasq$. With the improvement coefficients set to their
tree-level values (i.e. $\icsw = \ict = \icttil =1$),
the dominant discretisation effects are expected to be $\Ogsqa$
and $\Oasq$; with one-loop coefficients we have $\Ogqa$ and
$\Oasq$ errors etc.  These statements refer to the chiral limit
(away from which, we must also take into consideration counterterms
proportional to the quark mass). Since we are in the framework of
mass independent renormalisation, working in the chiral limit is
adequate for our purposes. We have performed numerical simulations in
two regimes:

\begin{enumerate}

\item What we call ``unimproved action results'' (or ``unimproved
case'' for short) consists in setting $\icsw = 0\,\, $. Moreover, we
set $\icttil = 1$, while the one-loop value\footnote{This is a choice of
convenience: it is important to know for renormalisation purposes (see
eq.~(\ref{eq:ssfLat}) below) the dependence of the Schr\"odinger
functional renormalised coupling $\gbar(1/L)$ on the bare coupling
$g_0$. This dependence  is known
non-perturbatively~\cite{SFcoupling,SFmassRGI} for the pure Yang-Mills
action with this $\ict$ value. In any case, the choice for $\ict$ has no
bearing on the order of leading lattice artifacts.}
(eq.~(\ref{eq:ctPT}) truncated to O($g_0^2$)) is used for
$\ict$. Since the action in the lattice bulk
is unimproved, the dominant discretisation effects ought to be $\Oa$.

\item  What we call ``improved action results'' (or ``improved
case'' for short) consists in using the Clover action with a
non-perturbative $\icsw$. The one-loop value from eqs.~(\ref{eq:ctPT}) and 
(\ref{eq:cttPT}) is used for $\ict$ and
$\icttil$ respectively. Thus the dominant
discretisation errors should be $\Ogqa$ and $\Oasq$. Since
the former only arise from perturbatively improved boundary
counterterms, while everything in the lattice bulk is fully improved,
it is reasonable to expect that correlation functions are mostly
affected by $\Oasq$ errors.  Numerical support for this expectation
has been presented in ref.~\cite{SFmassRGI}.

\end{enumerate}


\section{The step scaling function}
\label{sec:ssf}

SF renormalisation schemes are mass independent; i.e.
simulations can be performed in the chiral limit. The
renormalisation scale is set at the lattice IR cutoff (i.e. $\mu =
1/L$); the renormalised coupling $\gbar(1/L)$ and quark mass  $\mbar(1/L)$ are
then only functions of $L$. The SF renormalised
coupling has been defined in ref.~\cite{SFcoupling}. The
renormalised quark mass is~\cite{SFmassRGI}
\begin{gather}
\label{eq:renqmass}
\mbar (1/L) = \lim_{a \to 0} \,\, \ZA(g_0) \,\, \ZP^{-1}(g_0, L/a) \,\, m(g_0)
\end{gather}
where $m(g_0)$ is defined in eq.~(\ref{eq:masscurr}) and the
renormalisation condition for the pseudoscalar operator is
\begin{gather}
\label{eq:rencond}
\ZP(g_0,L/a) \frac{\fP(L/2)}{\sqrt{f_1}} \Bigg \vert_{m = 0} = c(\theta,a/L) \,\, ,
\end{gather}
with $c(\theta,a/L)$ such that at tree level $\ZP(0,L/a)=1$.
We will always impose eq.~(\ref{eq:rencond}) at
$\theta=0.5$~\rep{SFmassRGI,SFpt1}, and hence eliminate any explicit reference
to $\theta$ from now on. 
The axial current normalisation $\ZA(g_0)$,
being scale independent, has no effect on the renormalisation group 
running of the quark mass; thus it is of no immediate consequence to the present work.

Here we are interested in the step scaling function of the quark mass, which is
defined in the chiral limit $m(g_0)=0$, for a lattice of a given
resolution $L/a$ and at fixed renormalised coupling $\gbar^2(1/L) = u$, by
\begin{gather}
\label{eq:ssfLat}
\SigmaP(u,a/L) = \frac{\ZP(g_0,2L/a)}{\ZP(g_0,L/a)}
\Bigg \vert_{m=0,~\gbar^2(1/L) = u } \,\, .
\end{gather}
This quantity is finite in the continuum limit
\begin{gather}
\label{eq:ssfCont}
\sigmaP(u) = \lim_{a \to 0} \SigmaP(u,a/L) =
\frac{\mbar(1/L)}{\mbar(1/(2L))} \Bigg \vert_{\gbar^2(1/L) = u }
\,\, .
\end{gather}
The physical meaning of $\sigmaP$ follows from the RG equation obeyed by the renormalised quark mass
\begin{gather}
\label{eq:rgDiff}
\mu \,\, \frac{\partial \mbar(\mu)}{\partial\mu} =
\tau\big(\gbar(\mu)\big) \,\, \mbar(\mu)
\end{gather}
(recall that $\mu=1/L$).
Upon integration of this equation between scales $L^{-1}$ and $(2L)^{-1}$ we obtain
\begin{gather}
\label{eq:rgInt}
\sigmaP(u) = \exp \left\{ -\int_{\gbar(1/L)}^{\gbar(1/(2L))}
  \frac{\tau(g)}{\beta(g)} \,\dif g \right\} \, ,
\end{gather}
with $\beta(g)$ the Callan-Symanzik function. Thus, $\sigmaP$ is
closely related to  the quark mass anomalous dimension.

The lattice SSF $\SigmaP$ is not unique: it depends on the details of
the lattice regularisation (e.g. the type of lattice action chosen,
the level of $\Oa$ improvement etc.). Its continuum limit, however,
should be unique (i.e. universality should hold), unless lattice QCD,
or at least the specific regularisation implemented here,
exhibits some unexpected pathology. This is what the present
paper has set out to explore, in the spirit of refs.~\cite{berber,SFuniv}.


\section{Discretisation effects in perturbation theory}
\label{sec:ssfPT}

\begin{figure}[t]
\centering\psfig{file=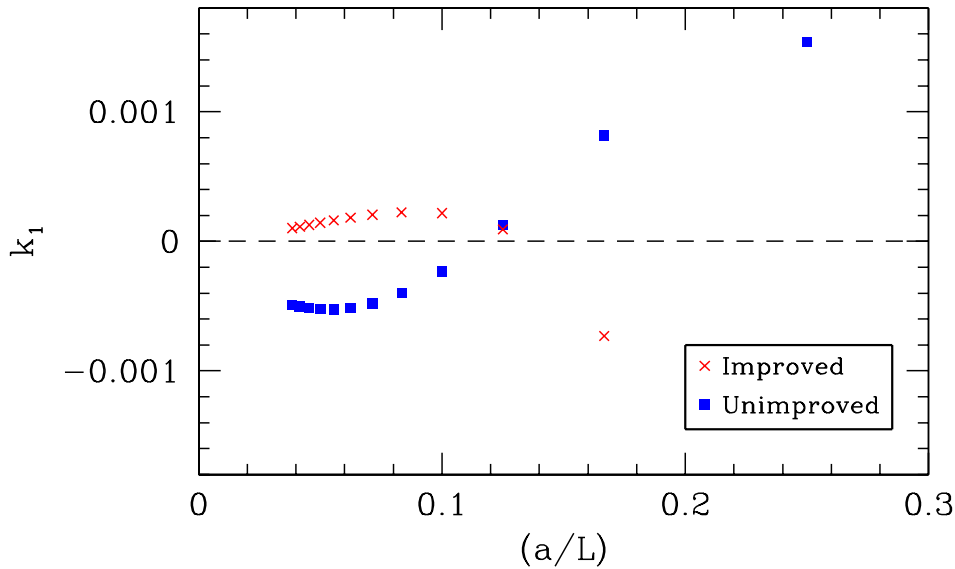,width=13.0cm}
\caption{
Cutoff dependence of the step scaling function
of the pseudoscalar density at one loop in perturbation theory.
}
\label{fig:k1}
\end{figure}

The expansion of the SSF in renormalised perturbation theory reads
\begin{gather}
\sigmaP(u) = 1 + \sum_{n=1}^\infty \sigmaP^{(n)} \, u^n \,\, ,
\end{gather}
with the LO universal RG coefficient $\sigmaP^{(1)} = - 8 \ln(2) / (4\pi)^2$.
In perturbation theory the cutoff dependence of the SSF can be studied by expanding
\begin{gather}
\frac{\SigmaP(u,a/L)}{\sigmaP(u)} = 1 + \sum_{n=0}^\infty k_n(a/L) \,\, u^n \,\, .
\label{eq:cutoffPT}
\end{gather}
Note that due to the choice of renormalisation
condition~(\ref{eq:rencond}), discretisation errors are absent at tree
level (i.e. $k_0(a/L) = 0$). Moreover, $k_n(0) = 0$ by construction.
The quantity $k_1(a/L)$, which contains the cutoff effects at one loop, is known 
for the improved case from the work of
Sint and Weisz~\cite{SFpt1} for $L/a= 4,6,\dots,16$. In their notation, it is given by
\begin{gather}
k_1(a/L) = k(\infty) \delta_k(L/a) =
-\frac{8\ln(2)}{(4 \pi)^2} \delta_k(L/a) \ ,
\end{gather}
with $\delta_k(L/a)$ tabulated in Table 2 of ref.~\cite{SFpt1} (the
case of interest to us is $\theta = 0.5,~\rho = T/L = 1$).
We have repeated these calculations for the unimproved case and extended
both cases to $L/a = 18, 20, \dots, 26$. The results are summarised in Table~\ref{tab:k1}
and Fig.~\ref{fig:k1}.

\begin{table}[t]
\centering
\begin{tabular}{c@{\hspace{10mm}}cc}
\Hline\\[-1.5ex]
$L/a$ & $k_1$ ({\rm Unimproved}) & $k_1$ ({\rm Improved}) \\
\hline\\[-1.0ex]
4  &  $\ \ 1.5366198\times10^{-3}$  &  $-7.4992105\times10^{-3}$ \\
6  &  $\ \ 8.1921362\times10^{-4}$  &  $-7.2993963\times10^{-4}$ \\
8  &  $\ \ 1.2463902\times10^{-4}$  &  $\ \ 9.2204726\times10^{-5}$ \\
10 &  $-2.3064991\times10^{-4}$     &  $\ \ 2.1695501\times10^{-4}$ \\
12 &  $-4.0199062\times10^{-4}$     &  $\ \ 2.2359743\times10^{-4}$ \\
14 &  $-4.8254066\times10^{-4}$     &  $\ \ 2.0536785\times10^{-4}$ \\
16 &  $-5.1712918\times10^{-4}$     &  $\ \ 1.8293858\times10^{-4}$ \\
18 &  $-5.2761682\times10^{-4}$     &  $\ \ 1.6171688\times10^{-4}$ \\
20 &  $-5.2516423\times10^{-4}$     &  $\ \ 1.4302219\times10^{-4}$ \\
22 &  $-5.1573079\times10^{-4}$     &  $\ \ 1.2694608\times10^{-4}$ \\
24 &  $-5.0261759\times10^{-4}$     &  $\ \ 1.1321941\times10^{-4}$ \\
26 &  $-4.8770374\times10^{-4}$     &  $\ \ 1.0149790\times10^{-4}$ \\[1.5ex]
\Hline
\end{tabular}
\caption{Results for the cutoff dependence of the step scaling function
of the pseudoscalar density at one loop in perturbation theory.}
\label{tab:k1}
\end{table}

The present perturbative analysis has been motivated by the wish to explore in detail
the cutoff dependence of our non-perturbative estimates of $\SigmaP(u,a/L)$, obtained
in our simulations at four lattice resolutions $L/a = 6,8,12,16$ (see next Section). 
Strictly speaking we therefore need to know $k_1(a/L)$ only at these four values of $(a/L)$,
for both the unimproved and improved cases. However it is clear that one-loop cutoff effects 
can have a rather non-trivial overall behaviour as $(a/L)$ is reduced. Had we limited our 
perturbative calculation to the range of interest (i.e.  $L/a =
6,8,12,16$), in the unimproved case we would have 
only observed that $k_1$ crosses over monotonically the abscissa axis at about $L/a = 8$, 
without signalling that it indeed
converges towards its limiting value $k_1(0) = 0$ (the improved
case already ``bends over'' towards the origin within this range). As
a partial safeguard against
the eventuality of some uncontrolled error afflicting our perturbative results (e.g. rounding 
in the numerical integrations), we have extended the calculations all the way to
$L/a = 26$. Table~\ref{tab:k1} and Fig.~\ref{fig:k1} demonstrate that indeed $k_1$ reaches a 
local extremum and subsequently points towards the origin of the axes, as it should.

In order to gain some further insight into this behaviour, we recall that the 
coefficients $k_n$ of the perturbative series can be expanded as
\begin{gather}
k_n(a/L) = \sum_{p=1}^\infty \left( \frac{a}{L} \right)^p \,\,
\sum_{l=0}^n c_{pl}^{(n)} \,\, [\, \ln(a/L) \, ]^l \,\, ,
\label{eq:delta_series}
\end{gather}
with the leading $\mbox{O}(u)$ (one-loop) discretisation effects having the form
\begin{gather}
\label{eq:Wcoeffs}
\begin{split}
k_1(a/L) =& \frac{a}{L} \bigg [ c_{10}^{(1)} + c_{11}^{(1)} \ln(a/L) \bigg] \\
     &+ \frac{a^2}{L^2} \bigg [ c_{20}^{(1)} + c_{21}^{(1)} \ln(a/L) \bigg]
+\mbox{O}(a^3/L^3) \,\, .
\end{split}
\end{gather}
Tree-level improvement implies in general that $c_{1n}^{(n)} = 0$; thus the $\mbox{O}(a/L)$ 
one-loop perturbative
contribution is as above, but without the logarithm (i.e. $c_{11}^{(1)} =  0$).
One-loop improvement implies that $c_{1n}^{(n)} = c_{1,n-1}^{(n)} = 0$; thus the $\mbox{O}(a/L)$
one-loop perturbative
contributions of eq.~(\ref{eq:Wcoeffs}) all vanish ($c_{10}^{(1)} = c_{11}^{(1)} = 0$).
The leading discretisation effects of $\SigmaP$ are then $\mbox{O}(u \, a^2/L^2)$.
Such ``dominant'' discretisation effects may in practice compete with the next order
$\mbox{O}(u^2 \, a/L)$ errors, arising from $k_2(a/L)$. In all cases the functional form of
$k_1(a/L)$ is clearly complicated; the observed behaviour can be explained by 
the strengths (and relative signs) of the various coefficients $c^{(n)}_{pl}$.

\section{Non-perturbative computation of the step scaling function}
\label{sec:ssfNP}

In this Section we study the extrapolation of $\SigmaP$ to the continuum limit.
We also obtain a very accurate estimate of the ratio of the RGI quark mass to its
renormalised counterpart at a hadronic scale.
The method of computation is identical to that of
ref.~\cite{SFmassRGI}.

\subsection{Continuum limit of the step scaling function}

For both the unimproved and improved cases the lattice SSF $\SigmaP$ has been evaluated at
14 values of the renormalised coupling $\gbar(1/L)$,
each for four lattice resolutions $L/a = 6,8,12$ and $16$.
Note that in ref.~\cite{SFmassRGI} only the improved case has been studied.
A full collection of our raw data is
presented in Tables~\ref{tab:data1} and~\ref{tab:data2}.
The ``tuning'' of $\beta$ at the four $L/a$ values, corresponding to an (almost) 
fixed renormalised coupling $\gbar^2(1/L) = u$ has been taken over from ref.~\cite{SFmassRGI}.
This corresponds to the first three columns of Tables~\ref{tab:data1} and~\ref{tab:data2}.
The same is true of $\hopc$ for the improved case (fourth column of Table~\ref{tab:data1}). All
other results are new\footnote{The computation of $\hopc$ for the unimproved case was first
performed in ref.~\cite{GJP} for 9 of the 14 couplings used here; following~\cite{SFmassRGI},
this is done at $\theta = 0$.}.
In the strong coupling regime new results have been obtained at $\gbar^2(1/L)=3.111$.
The statistical accuracy of our improved and unimproved $\SigmaP(u, a/L)$ results is comparable
(save for a few cases where the improved data has somewhat smaller
errors).\footnote{The typical statistics accumulated for small
  lattices is of several hundred configurations. For the largest
  lattices the number of configurations ranges from around 60 at the
  weakest couplings to around 200 at the strongest ones.}

A comparison of our data in the improved case with those of ref.~\cite{SFmassRGI} reveals
fairly compatible results: for $\ZP(g_0,L/a)$ and $\ZP(g_0,2L/a)$ we mostly
agree within errors, save for a few cases in which agreement is within $2\,\sigma$;
the same is true for $\SigmaP(u,a/L)$, with some data being compatible only within $1.5\,\sigma$.

\begin{table}[p]
\centering
\begin{tabular}{rrl@{\hspace{6mm}}llll}
\Hline\\[-1.5ex]
$\beta~~~$ & $\frac{L}{a}$ & $~~~\gbar^2(1/L)$ &
$~~~~~~\hopc$ & $\ZP\left(g_0,\frac{L}{a}\right)$ & $\ZP\left(g_0,\frac{2L}{a}\right)$ & $\SigmaP\left(u,\frac{a}{L}\right)$\\[1.0ex]
\hline \\[-1.0ex]
10.7503 & 6 & 0.8873(5) & 0.130591(4) & 0.8480(5) & 0.8192(8) & 0.9660(11)\\
11.0000 & 8 & 0.8873(10) & 0.130439(3) & 0.8402(5) & 0.8125(10) & 0.9670(13)\\
11.3384 & 12 & 0.8873(30) & 0.130251(2) & 0.8331(8) & 0.8049(11) & 0.9662(16)\\
11.5736 & 16 & 0.8873(25) & 0.130125(2) & 0.8253(8) & 0.7986(15) & 0.9676(20)\\
  [1.0ex]
10.0500 & 6 & 0.9944(7) & 0.131073(5) & 0.8326(5) & 0.8012(8) & 0.9623(11)\\
10.3000 & 8 & 0.9944(13) & 0.130889(3) & 0.8260(7) & 0.7957(9) & 0.9633(14)\\
10.6086 & 12 & 0.9944(30) & 0.130692(2) & 0.8153(8) & 0.7826(13) & 0.9599(19)\\
10.8910 & 16 & 0.9944(28) & 0.130515(2) & 0.8102(7) & 0.7796(15) & 0.9622(20)\\
  [1.0ex]
9.5030 & 6 & 1.0989(8) & 0.131514(5) & 0.8200(6) & 0.7831(10) & 0.9550(14)\\
9.7500 & 8 & 1.0989(13) & 0.131312(3) & 0.8117(6) & 0.7769(9) & 0.9571(13)\\
10.0577 & 12 & 1.0989(40) & 0.131079(3) & 0.8005(9) & 0.7668(11) & 0.9579(17)\\
10.3419 & 16 & 1.0989(44) & 0.130876(2) & 0.7959(10) & 0.7630(11) & 0.9587(18)\\
  [1.0ex]
8.8997 & 6 & 1.2430(13) & 0.132072(9) & 0.8013(4) & 0.7633(8) & 0.9526(11)\\
9.1544 & 8 & 1.2430(14) & 0.131838(4) & 0.7945(5) & 0.7548(11) & 0.9500(15)\\
9.5202 & 12 & 1.2430(35) & 0.131503(3) & 0.7842(7) & 0.7498(11) & 0.9561(16)\\
9.7350 & 16 & 1.2430(34) & 0.131335(3) & 0.7774(11) & 0.7407(14) & 0.9528(22)\\
  [1.0ex]
8.6129 & 6 & 1.3293(12) & 0.132380(6) & 0.7909(6) & 0.7501(11) & 0.9484(16)\\
8.8500 & 8 & 1.3293(21) & 0.132140(5) & 0.7826(7) & 0.7435(11) & 0.9500(16)\\
9.1859 & 12 & 1.3293(60) & 0.131814(3) & 0.7738(10) & 0.7348(15) & 0.9496(23)\\
9.4381 & 16 & 1.3293(40) & 0.131589(2) & 0.7661(9) & 0.7273(19) & 0.9494(27)\\
  [1.0ex]
8.3124 & 6 & 1.4300(20) & 0.132734(10) & 0.7808(5) & 0.7356(8) & 0.9421(12)\\
8.5598 & 8 & 1.4300(21) & 0.132453(5) & 0.7727(6) & 0.7282(11) & 0.9424(16)\\
8.9003 & 12 & 1.4300(50) & 0.132095(3) & 0.7621(10) & 0.7195(14) & 0.9441(22)\\
9.1415 & 16 & 1.4300(58) & 0.131855(3) & 0.7551(8) & 0.7129(16) & 0.9441(23)\\
  [1.0ex]
7.9993 & 6 & 1.5553(15) & 0.133118(7) & 0.7659(4) & 0.7178(12) & 0.9372(16)\\
8.2500 & 8 & 1.5553(24) & 0.132821(5) & 0.7575(7) & 0.7127(12) & 0.9409(18)\\
8.5985 & 12 & 1.5533(70) & 0.132427(3) & 0.7484(11) & 0.7021(14) & 0.9381(23)\\
8.8323 & 16 & 1.5533(70) & 0.132169(3) & 0.7405(11) & 0.6966(19) & 0.9407(29)\\
  [1.5ex]
\Hline
\end{tabular}
\caption{
Results for the step scaling function $\SigmaP$, improved case.
}
\label{tab:data1}
\addtocounter{table}{-1}
\end{table}

\begin{table}[p]
\centering
\begin{tabular}{rrl@{\hspace{6mm}}llll}
\Hline\\[-1.5ex]
$\beta~~~$ & $\frac{L}{a}$ & $~~~\gbar^2(1/L)$ &
$~~~~~~\hopc$ & $\ZP\left(g_0,\frac{L}{a}\right)$ & $\ZP\left(g_0,\frac{2L}{a}\right)$ & $\SigmaP\left(u,\frac{a}{L}\right)$\\[1.0ex]
\hline \\[-1.0ex]
7.7170 & 6 & 1.6950(26) & 0.133517(8) & 0.7527(6) & 0.6997(4) & 0.9296(9)\\
7.9741 & 8 & 1.6950(28) & 0.133179(5) & 0.7452(6) & 0.6934(11) & 0.9305(17)\\
8.3218 & 12 & 1.6950(79) & 0.132756(4) & 0.7353(4) & 0.6858(14) & 0.9327(20)\\
8.5479 & 16 & 1.6950(90) & 0.132485(3) & 0.7266(12) & 0.6792(16) & 0.9348(27)\\
  [1.0ex]
7.4082 & 6 & 1.8811(22) & 0.133961(8) & 0.7345(7) & 0.6773(5) & 0.9221(11)\\
7.6547 & 8 & 1.8811(28) & 0.133632(6) & 0.7259(7) & 0.6712(12) & 0.9246(19)\\
7.9993 & 12 & 1.8811(38) & 0.133159(4) & 0.7174(4) & 0.6630(13) & 0.9242(19)\\
8.2415 & 16 & 1.8811(99) & 0.132847(3) & 0.7132(16) & 0.6578(14) & 0.9223(29)\\
  [1.0ex]
7.1214 & 6 & 2.1000(39) & 0.134423(9) & 0.7149(7) & 0.6512(5) & 0.9109(11)\\
7.3632 & 8 & 2.1000(45) & 0.134088(6) & 0.7069(6) & 0.6452(13) & 0.9127(20)\\
7.6985 & 12 & 2.1000(80) & 0.133599(4) & 0.6976(4) & 0.6370(14) & 0.9131(21)\\
7.9560 & 16 & 2.100(11) & 0.133229(3) & 0.6904(12) & 0.6348(12) & 0.9195(24)\\
  [1.0ex]
6.7807 & 6 & 2.4484(37) & 0.134994(11) & 0.6874(8) & 0.6112(5) & 0.8891(13)\\
7.0197 & 8 & 2.4484(45) & 0.134639(7) & 0.6796(7) & 0.6079(14) & 0.8945(23)\\
7.3551 & 12 & 2.4484(80) & 0.134141(5) & 0.6711(5) & 0.5978(15) & 0.8908(23)\\
7.6101 & 16 & 2.448(17) & 0.133729(4) & 0.6664(12) & 0.5996(13) & 0.8998(25)\\
  [1.0ex]
6.5512 & 6 & 2.770(7) & 0.135327(12) & 0.6628(8) & 0.5775(4) & 0.8713(12)\\
6.7860 & 8 & 2.770(7) & 0.135056(8) & 0.6551(8) & 0.5753(14) & 0.8782(24)\\
7.1190 & 12 & 2.770(11) & 0.134513(5) & 0.6487(5) & 0.5704(10) & 0.8793(17)\\
7.3686 & 16 & 2.770(14) & 0.134114(3) & 0.6452(14) & 0.5672(15) & 0.8791(30)\\
  [1.0ex]
6.3665 & 6 & 3.111(4) & 0.135488(6) & 0.6395(9) & 0.5427(13) & 0.8486(24)\\
6.6100 & 8 & 3.111(6) & 0.135339(3) & 0.6356(8) & 0.5466(15) & 0.8600(26)\\
6.9322 & 12 & 3.111(12) & 0.134855(3) & 0.6290(12) & 0.5363(15) & 0.8526(29)\\
7.1911 & 16 & 3.111(16) & 0.134411(3) & 0.6286(9) & 0.5438(16) & 0.8651(28)\\
  [1.0ex]
6.2204 & 6 & 3.480(8) & 0.135470(15) & 0.6179(4) & 0.5058(12) & 0.8186(20)\\
6.4527 & 8 & 3.480(14) & 0.135543(9) & 0.6129(5) & 0.5085(17) & 0.8297(29)\\
6.7750 & 12 & 3.480(39) & 0.135121(5) & 0.6092(10) & 0.5102(15) & 0.8375(28)\\
7.0203 & 16 & 3.480(21) & 0.134707(4) & 0.6050(10) & 0.5056(17) & 0.8357(31)\\
  [1.5ex]
\Hline
\end{tabular}
\caption{
(continued)
}
\end{table}

\begin{table}[p]
\centering
\begin{tabular}{rrl@{\hspace{6mm}}llll}
\Hline\\[-1.5ex]
$\beta~~~$ & $\frac{L}{a}$ & $~~~\gbar^2(1/L)$ &
$~~~~~~\hopc$ & $\ZP\left(g_0,\frac{L}{a}\right)$ & $\ZP\left(g_0,\frac{2L}{a}\right)$ & $\SigmaP\left(u,\frac{a}{L}\right)$\\[1.0ex]
\hline \\[-1.0ex]
10.7503 & 6 & 0.8873(5) & 0.134696(7) & 0.8559(5) & 0.8290(7) & 0.9686(10) \\ 
11.0000 & 8 & 0.8873(10) & 0.134548(6) & 0.8450(5) & 0.8188(8) & 0.9690(11) \\ 
11.3384 & 12 & 0.8873(30) & 0.134277(5) & 0.8336(6) & 0.8066(10) & 0.9676(14) \\ 
11.5736 & 16 & 0.8873(25) & 0.134068(6) & 0.8264(7) & 0.8003(13) & 0.9684(18) \\ 
  [1.0ex]
10.0500 & 6 & 0.9944(7) & 0.135659(8) & 0.8413(5) & 0.8123(8) & 0.9655(11) \\ 
10.3000 & 8 & 0.9944(13) & 0.135457(5) & 0.8310(5) & 0.8012(9) & 0.9641(12) \\ 
10.6086 & 12 & 0.9944(30) & 0.135160(4) & 0.8188(7) & 0.7887(12) & 0.9632(17) \\ 
10.8910 & 16 & 0.9944(28) & 0.134849(6) & 0.8108(8) & 0.7826(16) & 0.9652(22) \\ 
  [1.0ex]
9.5030 & 6 & 1.0989(8) & 0.136520(5) & 0.8292(6) & 0.7973(8) & 0.9615(12) \\ 
9.7500 & 8 & 1.0989(13) & 0.136310(3) & 0.8189(5) & 0.7847(9) & 0.9582(12) \\ 
10.0577 & 12 & 1.0989(40) & 0.135949(4) & 0.8060(8) & 0.7739(11) & 0.9602(17) \\ 
10.3419 & 16 & 1.0989(44) & 0.135572(4) & 0.7980(12) & 0.7641(11) & 0.9575(20) \\ 
  [1.0ex]
8.8997 & 6 & 1.2430(13) & 0.137706(5) & 0.8119(6) & 0.7775(8) & 0.9576(12) \\ 
9.1544 & 8 & 1.2430(14) & 0.137400(4) & 0.8009(6) & 0.7651(9) & 0.9553(13) \\ 
9.5202 & 12 & 1.2430(35) & 0.136855(2) & 0.7880(8) & 0.7521(12) & 0.9544(18) \\ 
9.7350 & 16 & 1.2430(34) & 0.136523(4) & 0.7805(9) & 0.7452(14) & 0.9548(21) \\ 
  [1.0ex]
8.6129 & 6 & 1.3293(12) & 0.138346(6) & 0.8045(7) & 0.7654(8) & 0.9514(13) \\ 
8.8500 & 8 & 1.3293(21) & 0.138057(4) & 0.7912(6) & 0.7525(10) & 0.9511(15) \\ 
9.1859 & 12 & 1.3293(60) & 0.137503(2) & 0.7779(9) & 0.7378(12) & 0.9485(19) \\ 
9.4381 & 16 & 1.3293(40) & 0.137061(4) & 0.7703(13) & 0.7286(15) & 0.9459(25) \\ 
  [1.0ex]
8.3124 & 6 & 1.4300(20) & 0.139128(11) & 0.7905(7) & 0.7517(9) & 0.9509(14) \\ 
8.5598 & 8 & 1.4300(21) & 0.138742(7) & 0.7800(6) & 0.7377(11) & 0.9458(16) \\ 
8.9003 & 12 & 1.4300(50) & 0.138120(8) & 0.7669(10) & 0.7262(17) & 0.9469(25) \\ 
9.1415 & 16 & 1.4300(58) & 0.137655(5) & 0.7586(9) & 0.7190(17) & 0.9478(25) \\ 
  [1.0ex]
7.9993 & 6 & 1.5553(15) & 0.140003(11) & 0.7808(7) & 0.7350(9) & 0.9413(14) \\ 
8.2500 & 8 & 1.5553(24) & 0.139588(8) & 0.7671(6) & 0.7237(11) & 0.9434(16) \\ 
8.5985 & 12 & 1.5533(70) & 0.138847(6) & 0.7560(9) & 0.7083(16) & 0.9369(24) \\ 
8.8323 & 16 & 1.5533(70) & 0.138339(7) & 0.7458(13) & 0.6992(18) & 0.9375(29) \\ 
  [1.5ex]
\Hline
\end{tabular}
\caption{
Results for the step scaling function $\SigmaP$, unimproved case.
}
\label{tab:data2}
\addtocounter{table}{-1}
\end{table}

\begin{table}[p]
\centering
\begin{tabular}{rrl@{\hspace{6mm}}llll}
\Hline\\[-1.5ex]
$\beta~~~$ & $\frac{L}{a}$ & $~~~\gbar^2(1/L)$ &
$~~~~~~\hopc$ & $\ZP\left(g_0,\frac{L}{a}\right)$ & $\ZP\left(g_0,\frac{2L}{a}\right)$ & $\SigmaP\left(u,\frac{a}{L}\right)$\\[1.0ex]
\hline \\[-1.0ex]
7.7170 & 6 & 1.6950(26) & 0.140954(12) & 0.7650(7) & 0.7195(9) & 0.9405(15) \\ 
7.9741 & 8 & 1.6950(28) & 0.140438(8) & 0.7550(7) & 0.7095(15) & 0.9397(22) \\ 
8.3218 & 12 & 1.6950(79) & 0.139589(6) & 0.7418(10) & 0.6940(16) & 0.9356(25) \\ 
8.5479 & 16 & 1.6950(90) & 0.139058(6) & 0.7328(11) & 0.6823(19) & 0.9311(29) \\ 
  [1.0ex]
7.4082 & 6 & 1.8811(22) & 0.142145(11) & 0.7489(7) & 0.6994(10) & 0.9339(16) \\ 
7.6547 & 8 & 1.8811(28) & 0.141572(9) & 0.7368(7) & 0.6829(13) & 0.9268(20) \\ 
7.9993 & 12 & 1.8811(38) & 0.140597(6) & 0.7241(11) & 0.6725(15) & 0.9287(25) \\ 
8.2415 & 16 & 1.8811(99) & 0.139900(6) & 0.7161(12) & 0.6652(16) & 0.9289(27) \\ 
  [1.0ex]
7.1214 & 6 & 2.1000(39) & 0.143416(11) & 0.7309(8) & 0.6746(10) & 0.9230(17) \\ 
7.3632 & 8 & 2.1000(45) & 0.142749(9) & 0.7181(7) & 0.6564(17) & 0.9141(25) \\ 
7.6985 & 12 & 2.1000(80) & 0.141657(6) & 0.7037(8) & 0.6440(13) & 0.9152(21) \\ 
7.9560 & 16 & 2.100(11) & 0.140817(7) & 0.6980(12) & 0.6399(15) & 0.9168(27) \\ 
  [1.0ex]
6.7807 & 6 & 2.4484(37) & 0.145286(11) & 0.7057(8) & 0.6403(11) & 0.9073(19) \\ 
7.0197 & 8 & 2.4484(45) & 0.144454(7) & 0.6921(8) & 0.6224(12) & 0.8993(20) \\ 
7.3551 & 12 & 2.4484(80) & 0.143113(6) & 0.6796(8) & 0.6065(19) & 0.8924(30) \\ 
7.6101 & 16 & 2.448(17) & 0.142107(6) & 0.6745(12) & 0.6095(19) & 0.9036(32) \\ 
  [1.0ex]
6.5512 & 6 & 2.770(7) & 0.146825(11) & 0.6839(9) & 0.6083(11) & 0.8895(20) \\ 
6.7860 & 8 & 2.770(7) & 0.145859(7) & 0.6702(8) & 0.5938(17) & 0.8860(27) \\ 
7.1190 & 12 & 2.770(11) & 0.144299(8) & 0.6583(11) & 0.5796(14) & 0.8804(26) \\ 
7.3686 & 16 & 2.770(14) & 0.143175(7) & 0.6532(15) & 0.5772(19) & 0.8836(35) \\ 
  [1.0ex]
6.3665 & 6 & 3.111(4) & 0.148317(10) & 0.6635(9) & 0.5770(11) & 0.8696(20) \\ 
6.6100 & 8 & 3.111(6) & 0.147112(7) & 0.6529(9) & 0.5642(14) & 0.8641(25) \\ 
6.9322 & 12 & 3.111(12) & 0.145371(7) & 0.6394(11) & 0.5504(20) & 0.8608(35) \\ 
7.1911 & 16 & 3.111(16) & 0.144060(8) & 0.6329(13) & 0.5479(17) & 0.8657(32) \\ 
  [1.0ex]
6.2204 & 6 & 3.480(8) & 0.149685(15) & 0.6473(10) & 0.5466(13) & 0.8444(24) \\ 
6.4527 & 8 & 3.480(14) & 0.148391(9) & 0.6309(9) & 0.5315(23) & 0.8424(38) \\ 
6.7750 & 12 & 3.480(39) & 0.146408(7) & 0.6201(9) & 0.5218(21) & 0.8415(36) \\ 
7.0203 & 16 & 3.480(21) & 0.145025(8) & 0.6131(11) & 0.5177(20) & 0.8444(36) \\
  [1.5ex]
\Hline
\end{tabular}
\caption{
(continued)
}
\end{table}

If quenched lattice QCD has a universal continuum limit, then both sets of $\SigmaP$ results
(improved and unimproved action) ought to extrapolate to the same continuum value $\sigmaP$
at  fixed coupling $u$. What we have set out to investigate is the power dependence (linear
and/or quadratic) of the results on $(a/L)$. From the analysis of ref.~\cite{SFmassRGI} we
expect the dominant discretisation effects to be $\Oa$ in the unimproved case and $\Oasq$
in the improved one. Nevertheless we have performed fits on both datasets with the two Ans\"atze
\begin{align}
\label{eq:linfit}
\SigmaP (u,a/L) &= \sigmaP(u) + \rho(u)  (a/L) \ ,\\
\label{eq:quafit}
\SigmaP (u,a/L) &= \sigmaP(u) + \rho(u)  (a/L)^2 \ .
\end{align}
Another issue raised in ref.~\cite{SFmassRGI} is the number of data points which
should be included in each fit. In that work the $L/a=6$ results were dropped from
the fits, being too far from the continuum limit. We have performed fits with
all data (4-point fits) and also without the $L/a=6$ data (3-point fits). This
means that we have applied a total of four fitting procedures (the two Ans\"atze
of eqs.~(\ref{eq:linfit},\ref{eq:quafit}), each for a 3- and a 4-point fit).

\begin{figure}[p]
\centering
\hspace*{-3mm}\psfig{file=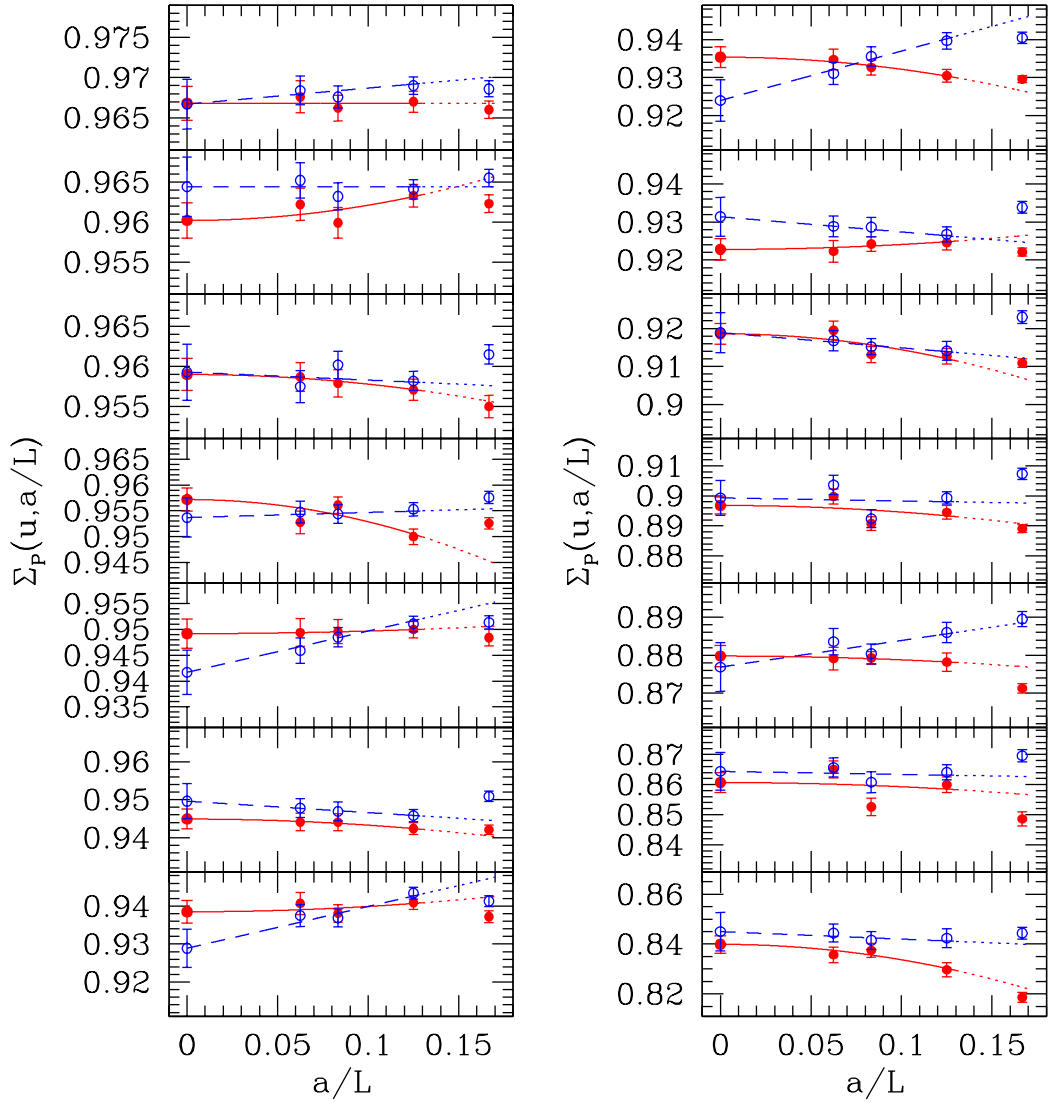,width=15.0cm}
\caption{
Continuum extrapolations of $\SigmaP$ at fixed renormalised coupling $u$
for the improved action (full symbols, solid line) and the unimproved action
(open symbols, dashed line).
The $L/a=6$ data points have not been included in the fits. 
The value of $u$ increases from top to bottom and from left to right.
}
\label{fig:extrap_P}
\end{figure}

The results of these fitting procedures can be summarised as follows:
\begin{enumerate}
\item In all cases, the statistical accuracy of our result for $\sigmaP$ is
better than $1\%$. The results for the linear or quadratic coefficients
$\rho$ have large statistical uncertainties (up to $100\%$), reflecting
an overall weak cutoff dependence of $\SigmaP$.
\item For any given lattice regularisation (i.e. improved or unimproved) and with any given
fitting Ansatz (i.e. linear or quadratic in $(a/L)$), the results for $\sigmaP$ obtained by
a 3-point fit are compatible to those obtained by a 4-point fit (at fixed coupling $u$).
Naturally, the former have a larger error.
\item For either lattice regularisation (i.e. improved or unimproved) and with any given
number of fitting points (i.e. 3-point fit or 4-point fit) the results for $\sigmaP$ 
obtained by a linear fit in $(a/L)$ are compatible to those obtained by a quadratic fit in $(a/L)$
(at fixed coupling $u$). There is just one exception for the improved data at the strongest
coupling $u=3.480$ with a 4-point fit (agreement is within $1.5 \sigma$). The results from
the quadratic fit are more accurate, due to the fact that the
extrapolation from the range of simulated data points to the continuum limit is shorter
in $(a/L)^2$ than in $(a/L)$.
\item The goodness of fit is always satisfactory ($\chi^2/{\rm d.o.f.} < 3$) at weak and intermediate
couplings ($ u \in [0.8873,1.8811]$). In a limited number of cases at stronger couplings the value
tends to rise considerably, but this apparently does not depend systematically on the number of
fitted points and choice of fitting Ansatz. In any case, given the small number of fitted data points,
$\chi^2/{\rm d.o.f.}$ is a goodness-of-fit criterion of relatively limited value. Instead, the 
total $\chi^2/{\rm d.o.f.}$ varies between 1 and 2, indicating satisfactory overall quality of the fits.
\end{enumerate}
We conservatively consider our 3-point fit results to be our best (i.e. we drop the data
computed at the largest lattice spacing) and opt for linear fits in $(a/L)$ with the 
unimproved case and quadratic ones with the improved one. The results for these options 
are shown in Fig.~\ref{fig:extrap_P}.

One could attempt to enrich this analysis along the lines of ref.~\cite{SFuniv}:
we recall that the discretisation effects known from perturbation
theory  (see eq.~(\ref{eq:cutoffPT}) and the related discussion) can be divided out of the lattice SSF,
by defining the quantity
\begin{gather}
\SigmaP^{(2)}(u,a/L) = \frac{\SigmaP(u,a/L)}{1 + u \,\, k_1(a/L) } \,\, .
\label{eq:Sigma2}
\end{gather}
The continuum limit of $\SigmaP^{(2)}$ is trivially the same as that of $\SigmaP$,
but the former quantity may approach it faster, as it has discretisation errors which are of 
order $u^2$. However we have seen in the previous Section that $k_1(a/L)$ is always numerically 
very small. Thus the denominator of eq.~(\ref{eq:Sigma2}) has an imperceptible impact on $\SigmaP$.

The continuum extrapolations of $\SigmaP$ (obtained with improved and unimproved lattice actions) 
give results which are fully compatible both in the weak and strong coupling regions.
At intermediate couplings we only have agreement within $1.5\,\sigma$; see Fig.~\ref{fig:extrap_P}.
The previous fitting analysis strongly suggests that this small discrepancy, rather than signalling
a lack of continuum limit universality, is to be attributed to discretisation effects not being
fully under control.

We will now corroborate this conclusion, by fitting our best results for the continuum 
SSF $\sigmaP(u)$ with the polynomial
\begin{gather}
\sigmaP(u) = 1 + \sum_{n=1}^N s_n u^n \,\,\, .
\label{eq:polyn}
\end{gather}
In all cases the first order coefficient is fixed to its PT value,
$s_1 = -8 \ln(2)/(4 \pi)^2$. One-parameter fits with $N=2$ yield
\begin{alignat}{3}
s_2 &= -0.0029(2) \qquad &(\chi^2/{\rm d.o.f.} \sim 1.0)
\qquad &{\rm improved~case} \ ,
\nonumber \\
s_2 &= -0.0028(3) \qquad &(\chi^2/{\rm d.o.f.} \sim 1.1)
\qquad &{\rm unimproved~case} \ ,
\label{eq:s2}
\end{alignat}
which are not too far from the PT value $s_2 = -0.002031(4)$ of ref.~\cite{SFpt1}.
One-parameter fits with $s_2$ fixed by PT and $N=3$ yield
\begin{alignat}{3}
s_3 &= -0.00031(5) \qquad &(\chi^2/{\rm d.o.f.} \sim 0.7)
\qquad &{\rm improved~case} \ ,
\nonumber \\
s_3 &= -0.00025(11) \qquad &(\chi^2/{\rm d.o.f.} \sim 1.1)
\qquad &{\rm unimproved~case} \ .
\label{eq:s3}
\end{alignat}
The above results are  compatible for the two lattice actions and thus
supportive of universality. This analysis becomes unstable once we push it
to two- or more-parameter fits. For instance, an $N=3$ fit with two fitting
parameters ($s_2$ and $s_3$) yields results with errors that range between 50\%
and 100\%, while the $N=4$ fits with either two ($s_3, s_4$)
or three fitting parameters ($s_2, s_3, s_4$) estimate them with 100\% uncertainty.

Having mustered adequate numerical support for universality, we
follow ref.~\cite{GJP} and calculate $\sigmaP(u)$ (at fixed coupling $u$)
by combined extrapolation of the $\SigmaP(u, a/L)$ data from both actions, constrained 
to a unique continuum limit. The improved (unimproved) case is assumed to depend
quadratically (linearly) on $a/L$. Results for $\sigmaP$ obtained with 3- and 4-point fits
are fully compatible at all couplings, while those for $\rho(u)$ are ill-determined, as
they carry up to 100\% uncertainties. The goodness of fit is mostly $\chi^2/{\rm d.o.f.}
\sim 1$, except for a couple of cases where it is around 4; anyway its average for all
couplings drops below 1.

\begin{figure}[p]
\centering
\vspace{15cm}
\includegraphics{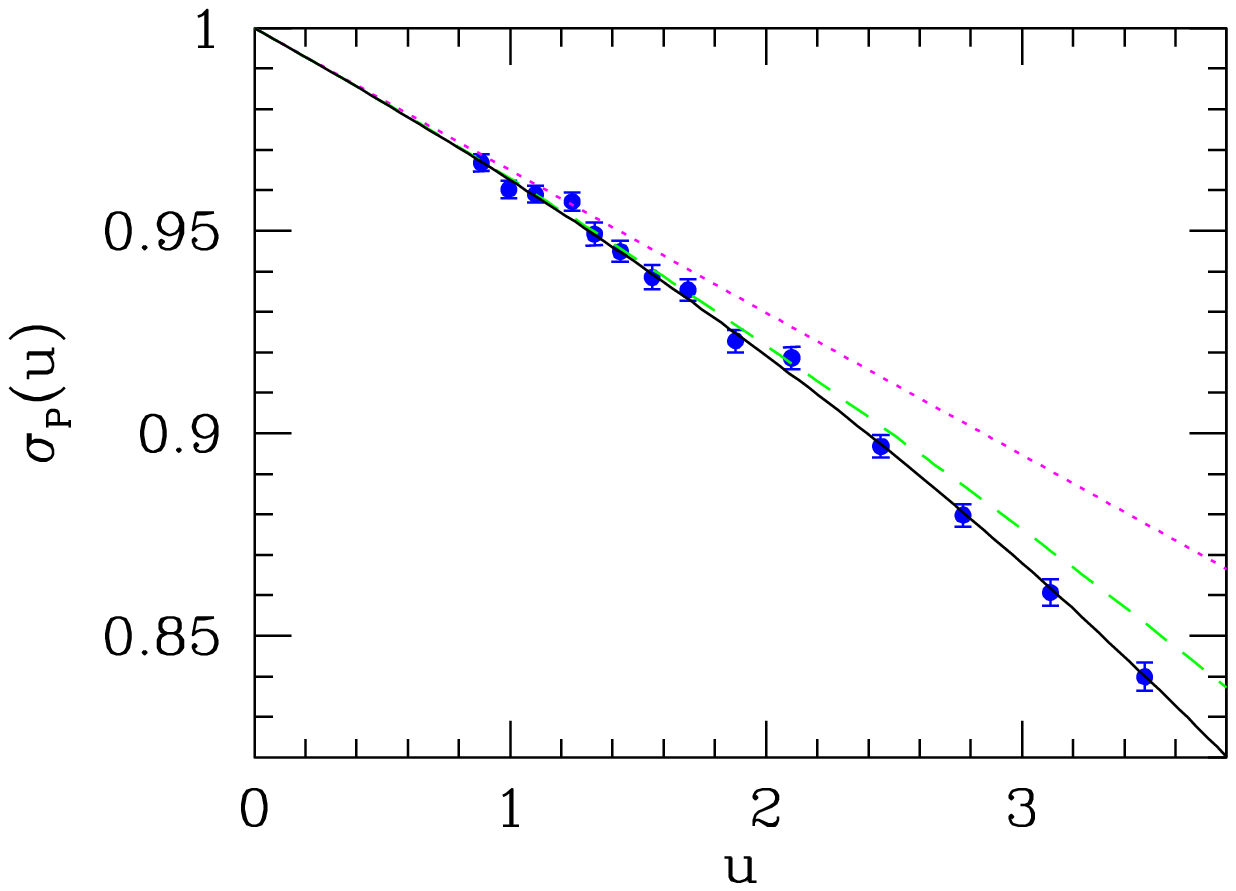}
\includegraphics{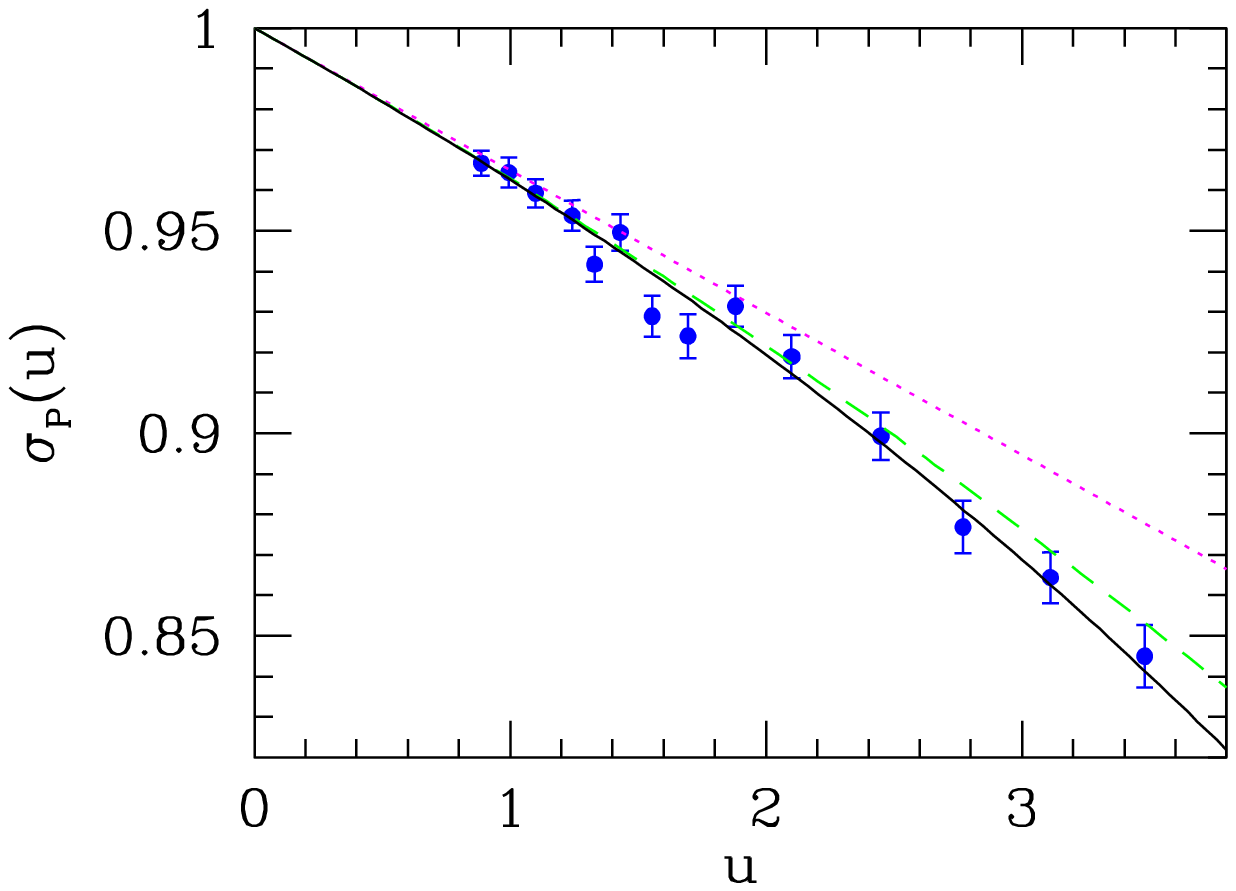}
\includegraphics{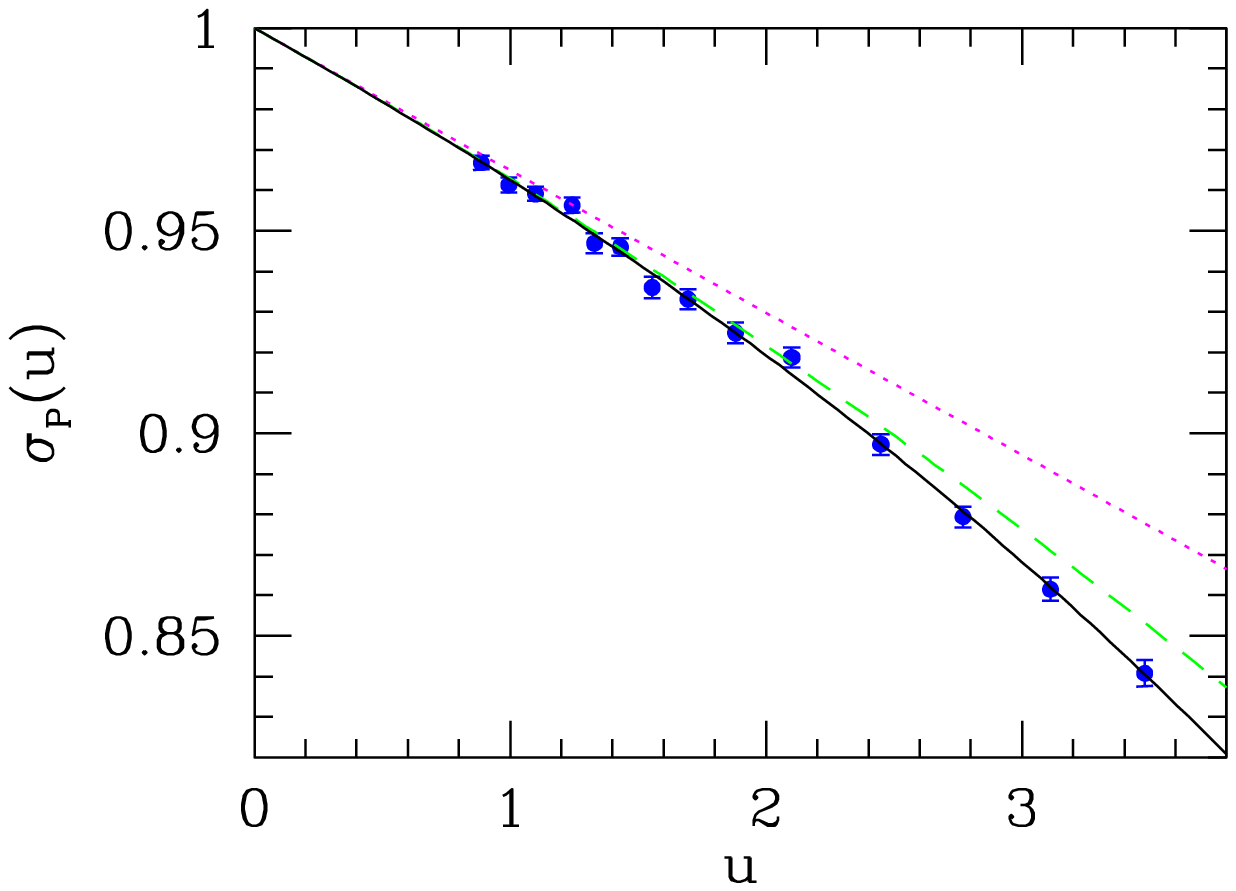}
\vspace{1.5cm}
\caption{
The step scaling function $\sigmaP(u)$ (full points) in the improved (top), unimproved
(middle) and combined (bottom) cases. Shown are also the expressions
for the step scaling function in LO (dotted line) and NLO (dashed line) perturbation
theory, as well as our best fit to the numerical data (solid line).
}
\label{fig:sigmaP}
\end{figure}

The 3-point fit results
for $\sigmaP(u)$ are subsequently fitted according to eq.~(\ref{eq:polyn}); with
$s_1$ given by PT, the case corresponding to eq.~(\ref{eq:s2}) gives
\begin{gather}
s_2 = -0.0028(1) \qquad (\chi^2/{\rm d.o.f.} \sim 1.1)
\qquad {\rm combined~case} \ ,
\end{gather}
while that of eq.~(\ref{eq:s3}) gives
\begin{gather}
s_3 = -0.00030(5) \qquad (\chi^2/{\rm d.o.f.} \sim 0.8)
\qquad {\rm combined~case} \ .
\label{eq:s3c}
\end{gather}
We take the results of eqs.~(\ref{eq:s3}) and (\ref{eq:s3c}) to be our best fits.
In Fig.~\ref{fig:sigmaP} we compare the LO and NLO predictions for the SSF with our
discrete non-perturbative data and the best-fit result. 

\subsection{RG running of the quark mass}

Using the functional form for $\sigmaP$ we can compute the ratio
of renormalised quark masses between  the minimum and maximum
renormalisation scales covered by our simulations. In order to
be consistent with the notation of ref.~\cite{SFmassRGI},
we denote the former by $(2\lmax)^{-1}$. The ratio in question is then
obtained in two steps:

First the SSF of the gauge coupling
\begin{gather}
\sigma(u)=\left.\gbar^2(1/2L)\right|_{\gbar^2(1/L)=u} \ ,
\end{gather}
computed in~\rep{SFcoupling, SFmassRGI},
is used in order to determine the correspondence between renormalised
couplings and renormalisation scales. This is done through the recursion
\begin{gather}
  \label{eq:ssf-u}
  u_{l}=\sigma(u_{l+1}) \, ,
\end{gather}
with $u_0 = \gbar^2(1/\lmax) = 3.48$ the initial value\footnote{This initial
value $u_0=3.48$ corresponds to $\lmax/r_0=0.738(16)$; the initial
calculation was performed in ref.~\cite{gsw} while the above result is quoted in
the more recent ref.~\cite{ns}.}. We note in passing that this procedure is based
on obtaining the SSF by fitting the results of refs.~\rep{SFcoupling, SFmassRGI}
by a polynomial
\begin{gather}
\sigma(u) = u \Big[ 1 + \sum_{n=1}^N \sigma_n u^n \Big] \,\,\, .
\label{eq:polyng}
\end{gather}
In the present analysis we have used the $N=4$ series, with $\sigma_1$, 
$\sigma_2$ fixed from PT and $\sigma_3$, $\sigma_4$ resulting from the fit.

Second the functional form for the SSF $\sigmaP$ is used for this sequence of couplings
in order to compute the mass ratio from the product (cf. eq.~(\ref{eq:ssfCont}))
\begin{gather}
  \label{eq:ratio}
  \frac{\mbar (1/2\lmax)}{\mbar (1/2^{-k+1}\lmax)} =
  \prod_{l=0}^{k-1}\left[\sigmaP(u_l)\right]^{-1} \, .
\end{gather}
In practice the range of scales covered by our simulations is spanned in $k=7$ iteration steps.

The final step in our calculation is the computation of the ratio of
the RGI quark mass $M$ to its scale dependent counterpart
$\mbar(\mu)$; in the quenched approximation this is given
by~\cite{SFmassRGI}
\begin{gather}
\frac{\mrgi}{\mbar(\mu)} = \bigg [ \frac{22}{(4 \pi)^2}\, \gbar^2(\mu) \bigg ]^{-4/11} 
\exp \left\{ - \int_0^{\gbar(\mu)} \dif g \left[ \frac{\tau(g)}{\beta(g)} 
    - \frac{8}{11 g} \right] \right\} \, .
\label{eq:Mrgi}
\end{gather}
In practice we compute the product of two ratios:
\begin{gather}
\frac{\mrgi}{\mbar(1/2\lmax)} =
\left(\frac{\mbar(1/2\lmax)}{\mbar(1/2^{-k+1}\lmax)}\right)^{-1}
\,\,\frac{\mrgi}{\mbar(1/2^{-k+1}\lmax)} \, .
\end{gather} 
The first ratio on the r.h.s. is known from eq.~(\ref{eq:ratio}). The
second ratio, which refers to a perturbative scale $\mu = 1/2^{-k+1}\lmax$,
is calculated from eq.~(\ref{eq:Mrgi}) with the NLO perturbative values of
$\beta(g)$ and $\tau(g)$. 

\begin{table}[t]
\centering
\begin{tabular}{c@{\hspace{5mm}}l@{\hspace{5mm}}l}
\Hline\\[-1.5ex]
{\rm Ref.} & {\rm Method} & $\relax{\kern-3pt}\frac{\mrgi}{\mbar(1/2\lmax)}\relax{\kern+3pt}$ \\[1.0ex]
\hline \\[-1.0ex]
\rep{SFmassRGI} & {\rm Improved}   & 1.157(12) \\
{\rm This work} & {\rm Improved}   & 1.154(9) \\
{\rm This work} & {\rm Unimproved} & 1.160(13) \\
{\rm This work} & {\rm Combined}   & 1.155(9) \\
  [1.5ex]
\Hline
\end{tabular}
\caption{Ratio of the RGI quark mass to the renormalised quark mass at scale 
$\mu = 1/2\lmax$. ``Method'' refers to the procedure used in the
computation of $\sigmaP$.}
\label{tab:Mrgi}
\end{table}
Having described the method, we gather the relevant results (and that of ref.~\cite{SFmassRGI})
in Table~\ref{tab:Mrgi}. The errors have been computed as outlined in 
Appendix B of ref.~\rep{SFmassRGI}. The following comments are in place:
\begin{enumerate}
\item
The quoted results have been obtained from the best SSF fits of eqs.~(\ref{eq:s3})
and (\ref{eq:s3c}). Several other fits, such as 
those  described in the previous subsection, have been tried out. In all cases the
final result $\mrgi/\mbar(1/2\lmax)$ fluctuated within the quoted error, which in turn
only increased slightly with increasing number of fitted parameters.
\item
Our improved result is compatible with that of ref.~\cite{SFmassRGI}.
The error is now smaller, due to improved statistics for the raw data on $\ZP$. 
\item
Compared to our unimproved result, the improved one has a smaller error.
Recalling that statistics are roughly the same, this reflects a better systematic
control of discretisation effects, such as stable quadratic extrapolations in $(a/L)$.
\item
The result of the combined case is identical to the improved one.
\end{enumerate}
Our final result is
\begin{gather}
  \frac{\mrgi}{\mbar(1/2\lmax)} = 1.155(9) \, .
\end{gather}
The quoted error does not include the effect of the
uncertainty in the determination of $\lmax/r_0$,
reported in ref.~\cite{ns}. Its contribution being roughly equal to
the above error implies that there is no point in increasing the
precision of our result unless the uncertainty in $\lmax/r_0$ is also
reduced.

\section{Conclusions}

We have performed a very detailed computation of the step scaling function of the quark mass
in quenched lattice QCD, employing two variants of the lattice regularisation,
namely unimproved and Clover-improved Wilson actions. In both cases the SSF has been computed
at many renormalised gauge couplings (corresponding to a wide range of renormalisation scales)
and for several lattice resolutions. Upon extrapolation to the continuum, the SSF has turned out
to be independent of the specifics of the lattice regularisation, providing convincing
evidence for the universality of the continuum limit.

The uniqueness of the continuum SSF has subsequently been used as a constraint, giving us
an extra handle for the control of the sensitive extrapolations to zero lattice spacing.
The final outcome of this detailed analysis, applied to high statistics data, is a very precise
value of the ratio $M/\mbar(\mu)$. Far from being an academic exercise, the increased 
accuracy of our result, compared to \cite{SFmassRGI}, is of practical relevance.
For example, in the context of the non-perturbative matching of Heavy Quark Effective Theory
and QCD in finite volume, recently proposed and applied in
refs.~\cite{mbstat},  precise numerical knowledge of
the functional dependence of QCD observables on the renormalisation group
invariant quark mass is of great importance.

The analysis described in this work is currently being applied to the SSF of other 
phenomenologically interesting quantities. Preliminary results on the
SSF of the tensor bilinear operator $\bar \psi \sigma_{0k} \psi$
(relevant e.g. to some semileptonic $B$-meson decays)
have appeared in ref.~\cite{Jpena}, while the first results on the SSF of
four-fermion operators (related to neutral meson oscillations, Kaon decays etc.) can be found in
ref.~\cite{ssf4f}.


\section*{Acknowledgments}
We thank P.~Hasenfratz, F.~Niedermayer and S.~Sint for discussions.
Special thanks go to R.~Sommer and H.~Wittig for a critical reading of the
manuscript and many helpful suggestions. C.P. acknowledges
the financial support provided through the European Community's Human
Potential Programme under contract HPRN-CT-2000-00145, Hadrons/Lattice
QCD. A.V. thanks the Bern Theory Group for its hospitality during the
initial stages of this work.


\clearpage



\end{document}